\documentclass[intlimits,twoside,a4paper]{article}

\usepackage[cp1251]{inputenc}

\usepackage[eqsecnum]{cmpj3}

\issue{2025}{28}{3}{33602}
\doinumber{10.5488/CMP.28.33602}

\title[The self-assembly behavior of a diblock copolymer/homopolymer induced by Janus nanorods]
{The self-assembly behavior of a diblock copolymer/homopolymer induced by Janus nanorods}

\author[Y. Q. Guo, J. Liu, H. R. He, N. Wu, J. J. Zhang]{Y. Q. Guo\orcid{0000-0003-1576-5148} \refaddr{label1}, J. Liu\orcid{0009-0008-3817-968X} 
\refaddr{label2,label3}, H. R. He\refaddr{label4}, N. Wu\refaddr{label1}, J. J. Zhang\refaddr{label2,label3}\thanks{Corresponding author: \email{zhangjinjun@sxnu.edu.cn}.}
}
\addresses{
\addr{label1} Department of Chemical Engineering and Materials Engineering, Lyuliang University, 033001 Lishi, China
\addr{label2} School of Physics and Information Engineering, Shanxi Normal University, 030031 Taiyuan, China
\addr{label3}Shanxi Institute of Coal Chemistry, Chinese Academy of Sciences, 033001 Taiyuan, China
\addr{label4} School of Chemistry and Chemical Engineering, Shanxi University, 030006 Taiyuan, China
}

\date{Received December 30, 2024, in final form April 11, 2025}
\begin{document}

\maketitle

\begin{abstract}
We employ cell dynamics simulation based on the CH/BD model to investigate the self-assembly behavior of a mixed system consisting of diblock copolymers (AB), homopolymers (C), and Janus nanorods. The results indicate that, at different component ratios, the mixed system undergoes various phase transitions with an increasing number of nanorods. Specifically, when the homopolymer component is 0.40, the mixed system transitions from a disordered structure to a parallel lamellar structure, subsequently to a tilted layered structure, and ultimately to a perpendicular lamellar structure as the number of nanorods increases. To explore this phenomenon in greater depth, we conduct a comprehensive analysis of domain sizes and pattern evolution. Additionally, we investigate the effects of the repulsive interaction strength between polymers, wetting strength, length of nanorods, and degree of asymmetry on the self-assembly behavior of the mixed system. This research provides significant theoretical and experimental insights for the preparation of novel nanomaterials.
%

%
\printkeywords
\end{abstract}

\section{Introduction}

Soft matter is characterized as a complex material state of matter that exists between solid and ideal fluid forms, distinguished by its unique self-assembly capabilities that facilitate the formation of intricate structures. This material exhibits pronounced macroscopic effects in response to minimal external stimuli, thereby demonstrating a wide range of physical and chemical properties  \cite{key1,key2,key3,key4}. In the field of materials science, a significant enhancement of material properties can be achieved through the blending of various types of atoms, compounds, or macromolecules, especially pronounced in polymer systems  \cite{key5}. Polymers, and block copolymers in particular, have demonstrated considerable potential to be applied in the fields such as nanolithography, optoelectronic devices, and biomedicine, owing to their unique molecular architectures and characteristics of microphase separation  \cite{key6,key7,key8,key9,key10,key11,key12,key13,key14,key15,key16}. Furthermore, polymer nanocomposites, which are synthesized by incorporating nanorods or nanoparticles into polymer matrices, exhibit remarkable properties in various fields such as optics, electronics, magnetics, and mechanics, thus attracting significant interest as a focal point of research within materials science  \cite{key17,key18,key19,key20,key21,key22,key23}.

Wu and Lu \cite{key24} explored the selective partitioning of silver nanowires in polystyrene/poly(vinyl pyrrolidone) (PS/PVP) blend films, demonstrating that the incorporation and surface modification of silver nanowires considerably affect the morphology of the blend, thereby offering potential applications in the regulation of phase-separated structures. Chiu et al. investigated the distribution and self-assembly behavior of gold nanoparticles in various block copolymers including polystyrene-poly(2-vinylpyridine) (PS-PVP) \cite{key25}, polystyrene-b-2-vinylpyridine (PS-P2VP) \cite{key26}, and polystyrene-b-poly(2-vinylpyridine) (PS-b-P2VP) \cite{key27,key28}. Their findings indicated that modifications to the surface chemistry of the nanoparticles facilitated a precise positioning and a controlled distribution within the block copolymer matrix. Yeh et al.~\cite{key29} investigated the morphological transitions induced by cadmium sulfide (CdS) nanoparticles in poly(styrene-b-4-vinylpyridine) (S4VP) block copolymers, revealing that the introduction of CdS nanoparticles led to a transformation of the S4VP block from a hexagonal close-packed cylindrical structure to a lamellar structure. Park et al. \cite{key30} studied the morphological transition in a bilayer film of poly (acrylic acid) (PAA) and polystyrene (PS) block copolymer (BCP) induced by magnetic nanoparticles (MNPs). Their research demonstrated that the aggregation of nanoparticles could induce a morphological transition in the BCP bilayer film, resulting in a shift from a bilayer structure to spherical micelles. Numerous additional studies have elucidated the mechanisms by which nanoparticles regulate the phase behavior of copolymers \cite{key31,key32,key33}. Li et al.~\cite{key34} synthesized a substantial quantity of ZnO nanorods utilizing polar polymer polyvinyl alcohol (PVA) as soft template,controlling the growth of the nanorods by varying the annealing temperature. Zhang et al.~\cite{key35} developed pH-responsive nanoparticle/polymer composite microcapsules through the polyamine-salt polymerization method. These composite microcapsules, which possess pH-sensing capabilities and drug-loading functionalities, are particularly advantageous for real-time monitoring of local intracellular pH during drug delivery processes.

In addition to experimental investigations, theoretical research concerning polymer nanocomposites has attracted considerable scholarly interest, with the anticipation that theoretical results will corroborate experimental findings and offer insights for the future developments of innovative nanomaterials. Balazs and his colleagues developed a coarse-grained model \cite{key36} to study the phase separation dynamics of nanoparticles and nanorods in a binary mixture. Their finding indicated that when nanoscale low-volume rods are immersed in a binary phase-separated blend, they self-assemble into a needle-like permeable network \cite{key37}, which confirmed experimental findings of Wu and Lu  \cite{key24}. Subsequently, they explored the self-assembly behavior of composites made from block copolymers and nanoparticles \cite{key38} and under confinement between two hard walls \cite{key39}, revealing the mechanism by which nanoparticles regulate the phase behavior of copolymers. Ma and his research team revealed the physical mechanisms of interactions between nanoparticles by using dissipative particle dynamics (DPD) methods \cite{key40} and studied the migration processes of nanoparticles \cite{key41}. They also researched on the morphological control of nanoparticles using Monte Carlo simulation methods \cite{key42}. Kalra et al.~\cite{key43} investigated the dispersion behavior of selective and non-selective nanoparticles in symmetric diblock copolymers under shear flow using molecular dynamics simulations. There findings indicated that shear flow has a substantial impact on the positioning of nanoparticles, which can serve as a parameter to control the self-assembly of nanocomposites. Selective nanoparticles exhibited a tendency to aggregate at the ends of the copolymer chains to which they are attracted, whereas non-selective nanoparticles predominantly localized at the interface between the two phases. Zhang and colleagues employed a third-order parameter model, examined
the phase behavior of diblock copolymer-homopolymer mixtures with modulated wettable particles~\cite{key44} and oscillatory particles~\cite{key45}. They subsequently analyzed the phase behavior in the binary blend of diblock copolymers mixtures induced by the oscillatory nanoparticles~\cite{key46}, and wettable nanoparticles \cite{key47} using cell dynamics simulation. Additionally, they investigated the self-assembly behavior of diblock copolymers~\cite{key18,key19} and block copolymer-homopolymer mixtures \cite{key48} induced by nanorods, thereby providing valuable insights for the development of novel high-performance nanomaterials. Diaz and colleagues used cell dynamics and brownian dynamics simulations to study the co-assembly behavior of nanoparticles in BCP systems, considering the effects of nanoparticles with different affinities, concentrations, and sizes on the phase behavior of BCPs \cite{key49,key50,key51,key52}. They also found that non-spherical nanoparticles can interact anisotropically with both the surrounding medium and with themselves \cite{key53}. Taylor and colleagues investigated the diffusion of thin nanorods within entangled polymer melts by applying coarse-grained molecular dynamics simulations, emphasizing the impact of nanorod length and roughness on their dynamics \cite{key54}. Zhou et al. \cite{key55} investigated the influence of nanorod surface properties on the phase separation kinetics and morphological transition of immiscible polymer blends under both shear and shear-free conditions through dissipative particle dynamics simulations. Osipov et al. \cite{key56} explored the orientational ordering and spatial distribution of nanorods of varying lengths within the lamellar phase of diblock copolymers using a combination of theoretical models and dissipative dynamics simulations. Huang also used this methodology to simulate and investigate the self-assembly behavior of mixtures of amphiphilic block copolymers and nanoparticles within nanotubes of various sizes and surface properties~\cite{key57}. This study demonstrated that by altering the dimensions and surface properties of the nanotubes, nanoparticles can be assembled into various structures, including nanowires, nanotubes, stacked disks, single helices, and double helices. Yao et al.~\cite{key58} examined the supramolecular polymerization behavior of polymeric nanorods mediated by block copolymers, observing that block copolymers  can modulate the interactions between nanorods, facilitating the formation of bundles. Furthermore, the BCPs ultimately led to the development of helical nanopatterns on the surfaces of these bundles. 

Janus nanoparticles, a new type of nanomaterial characterized by their unique asymmetry, outperform conventional nanoparticles due to their special structures and properties \cite{key59}. This lends them considerable research importance across various fields \cite{key59,key60,key61,key62,key63,key64,key65,key66}, consequently urging extensive interest and attention \cite{key67,key68,key69,key70,key71,key72,key73,key74,key75}. As one of the anisotropic particles, Janus nanorods are gaining significance as additives in the production of polymer composites. The incorporation of nanorods into polymeric materials results in composites that exhibit superior mechanical properties compared to polymers containing an equivalent volume fraction of spherical inclusions \cite{key18}. Yan and his colleagues demonstrated that the Cahn-Hilliard/Brownian dynamics (CH/BD) model is capable of effectively studying and predicting the self-assembly behavior of Janus nanorods in a polymer matrix \cite{key76}, as well as  regulating these behaviors through external conditions such as shear fields \cite{key77}. Subsequently, they employed dissipative particle dynamics to investigate the self-assembly behavior of nanoparticles at interfaces, highlighting the significant influence of entropy in the organization of nanoparticles \cite{key78,key79}. Diaz and his team utilized cell dynamics and Brownian dynamics simulation to investigate the co-assembly behavior of Janus nanoparticles within block copolymer systems \cite{key69}. Li et al.~\cite{key70} employed DPD method to investigate the role of Janus nanoparticles in the phase separation dynamics of polymer blends. Their findings showed that these nanoparticles have a dual function of accelerating decomposition at early stages and delaying it at later stages. Paiva et al. used DPD method to study the oriented assembly of Janus nanorods, demonstrating the tunability of their rolling behavior at interfaces through shear rate, interaction potential, and particle concentration~\cite{key80}. Zhou et al. also employed the DPD method to explore the impact of Janus nanorods on interfacial tension in A/B homopolymer blends, highlighting the significance of the rod length in controlling their orientation due to complex entropic and enthalpic interactions \cite{key81}. Wang et al. found that Janus nanoparticles placed at block copolymer interfaces through molecular dynamics~(MD) simulation can considerably improve the stress-strain behavior \cite{key82}. Burgos-Marmol et al. conducted a study on the structural and dynamic properties of Janus nanodimers dispersed in diblock copolymer lamellar phases using MD simulation, demonstrating that tailored interactions can facilitate a precise control over the spatial distribution and orientation of these nanodimers \cite{key83}. Osipov et al. also employed MD simulations to study the orientational ordering and spatial distribution of Janus nanoparticles in lamellar diblock copolymers, confirming their preferential localization and highly ordered arrangement at the boundary region, which is driven by differential affinities for distinct monomer structural domains \cite{key84}. These studies highlight the adaptability and potential applications of Janus nanoparticles and nanorods in modulating the properties of polymer systems.

In terms of experimentation, Liu et al. fabricated Janus particles of block copolymer/homopolymer (BCP/hP) blends through three-dimensional confined self-assembly (3DCSA) guided by dynamic neutral interfaces, and then employed a self-templating selective direct carbonization strategy to prepare composite asymmetric mesoporous carbon microparticles (MCMPs) \cite{key75}. Zhang et al. investigated the mechanisms governing the self-assembly of block copolymers within emulsion droplets, and prepared Janus particles through a combination of selective crosslinking and disassembly techniques \cite{key85}. Han et al. successfully synthesized Janus-structured nanorods harnessing hydrogen bonding interactions, demonstrating their potential in nanomaterial fabrication \cite{key86}. Li et al. successfully synthesized 3-miktoarm star terpolymers (PEG-star-PCL-star-P(CL-co-THF)) via Janus polymerization, yielding high-aspect-ratio nanorods and needle-like structures upon self-assembly at interfaces \cite{key87}. Yang et al. delved into how designing and manufacturing Janus nanoparticles can regulate their interfacial distribution within block copolymers, thereby influencing the structure and properties of the composites \cite{key88}. These studies not only highlight the critical significance of Janus nanoparticles in composite materials but also present the strategies for optimizing the properties of these composites through modifications in nanoparticle composition and processing parameters.

Our research focuses on thoroughly investigating the self-assembly behavior of a diblock copolymer/homopolymer/Janus nanorod mixed system, with the goal of developing nanomaterials that possess unique functions and advancing scientific research and technological innovation in related areas. Section~\ref{sec_2} outlines the computational methods used in this study, while section~\ref{sec_3} presents the results and discussions, and section~\ref{sec_4} concludes the findings.

\section{Theoretical models and simulation methods}
\label{sec_2}

We employ a combined method based on CH/BD (Cahn-Hilliard/Brownian Dynamics) model \cite{key36,key89,key90,key91} to investigate the self-assembly behavior of a symmetric diblock copolymer (AB)/homopolymer (C)/Janus nanorods mixed system. The diblock copolymer consists of A and B blocks connected by covalent bonds, while the homopolymer consists of iterating units of monomer C. In our model, we use the CH model to describe the polymer system, while for the Janus nanorods system, we use the BD model to describe it.

The dynamic equations for the diblock copolymer/homopolymer mixed system are complex, and their specific expressions are given as follows:
\begin{equation}
\frac{\partial\eta}{\partial t}=M_\eta\nabla^2 \frac{\delta F}{\delta\eta},
\label{eq:1}
\end{equation}
\begin{equation}
\frac{\partial\psi}{\partial t}=M_\psi\nabla^2 \frac{\delta F}{\delta\psi}.
\label{eq:2}
\end{equation}
In these equations, $M_\eta$ and $M_\psi$ denote the mobilities of the respective polymers. The variable $\psi=\phi_\text{A}-\phi_\text{B}$ describes phase separation in the diblock copolymer, while $\eta=\phi_\text{A}+\phi_\text{B}-\psi_\text{C}$ characterizes phase separation between the diblock copolymer and the homopolymer. The term $\psi_\text{C}={\sqrt{N_\text{C}}}/{(\sqrt{N_\text{AB}}+\sqrt{N_\text{C}})}$ represents the critical volume fraction for phase separation, which depends on the degree of polymerization of each component. Here, ${N_\text{AB}=N_\text{A}+N_\text{B}}$, where $N_\text{A}$, $N_\text{B}$, $N_\text{C}$ are the polymerization indices of blocks A, B, and hompolymer C, respectively.

The  motion of Janus nanorods is described by Langevin equations:
\begin{equation}
\frac{\partial \boldsymbol{r}_i}{\rd t}=-\frac{M_1\partial F}{\partial \boldsymbol{r}_i}+\zeta_i,
\label{eq:3}
\end{equation}
\begin{equation}
\frac{\partial\theta_i}{\rd t}=-\frac{M_2\partial F}{\partial \theta_i}+\xi_i,
\label{eq:4}
\end{equation} $M_1$ and $M_2$ are the mobility coefficients related to the motion and rotation of the nanorods, while $\zeta_i$ and $\xi_i$ denote thermal fluctuations
where $\boldsymbol{r}_i$ and $\theta_i$ represent the center-of-mass position and orientation angle of the $i$-th Janus nanorod, respectively. The parameters satisfy the fluctuation-dissipation relations.

The free energy $F$ \cite{key36,key92} of the diblock copolymer (AB)/homopolymer (C) /Janus nanorods mixed system is composed of three parts:
\begin{equation}
F=F_\text{GL}+F_\text{CPL}+F_\text{RR}.
\label{eq:F}
\end{equation}

$F_\text{GL}$ describes the Ginzburg-Landau free energy of the diblock copolymer:
\begin{align}
F_\text{GL}(\eta,\psi) &= \int \rd\boldsymbol{r}\left[ \left( -\frac{a}{2}\psi^2 + \frac{b}{4}\psi^4 + \frac{c}{2}(\nabla\psi)^2 \right) + \left( -\frac{a'}{2}\eta^2 + \frac{b'}{4}\eta^4 + \frac{c'}{2}(\nabla\eta)^2 \right) \right. \nonumber \\
& \left. +\, b_1\eta\psi - \frac{1}{2}b_2\eta\psi^2 \right] + \frac{\alpha}{2} \iint \rd\boldsymbol{r} \rd\boldsymbol{r}' G(\boldsymbol{r},\boldsymbol{r}') [\psi(\boldsymbol{r})-\psi_0][\psi(\boldsymbol{r}')-\psi_0] ,
\label{eq:F_GL} 
\end{align}
where $a$, $b$, $c$, $a'$, $b'$, $c'$, $b_1$ and $b_2$ are constants, $b_1$ \cite{key93,key94} is the repulsive interaction strength between polymers, expressed as $b_1={(\chi_\text{BC}-\chi_\text{AC})}/{2}$, $b_2$ \cite{key95} is defined as the expression $b_2={1}/{\psi_\text{C}N_\text{A}}$. $G(\boldsymbol{r},\boldsymbol{r}')$ is the Green's function defined by the equation $-\nabla^2G(\boldsymbol{r},\boldsymbol{r}')=\delta(\boldsymbol{r}-\boldsymbol{r}')$, $\alpha$ describes the strength of the long-range force owing to the covalent linkage between the A and B components. For a symmetric diblock copolymer, the average value of $\psi$ is denoted by $\psi_0=0$; by contrast, for asymmetric diblock copolymer, $\psi_0 \neq 0$ \cite{key96,key97,key98}.

$F_\text{CPL}$ describes the interaction free energy between the polymer and Janus nanorods, expressed as follows:
\begin{equation}
F_\text{CPL}=\int \rd\boldsymbol{r} \sum_{i}
\int \rd \boldsymbol{s}_iV_0\exp\left(\frac{-|\boldsymbol{r}-\boldsymbol{s}_i|}{r_0}\right)\big[W(\boldsymbol{r})-W_w\big]^2,
\label{eq:F_CPL}
\end{equation}
where $\boldsymbol{s}_i=\boldsymbol{r}_i+\delta \boldsymbol{s}_i$ denotes a point on the surface of the $i$-th Janus nanorod, and $\int\rd\boldsymbol{r}$ represents the integral over the length of the $i$-th nanorod. When $W(\boldsymbol{r})=\psi=\phi_\text{A}-\phi_\text{B}$ and $W_w=1$, Janus nanorods are preferentially wetted by phase A. When $W(\boldsymbol{r})=\psi=\phi_\text{A}-\phi_\text{B}$ and $W_w=-1$, Janus nanorods are preferentially wetted by phase B. When $W(\boldsymbol{r})=\eta=\phi_\text{A}+\phi_\text{B}-\psi_\text{C}$ and $W_w=-1$, Janus nanorods are preferentially wetted by phase C. $V(\boldsymbol{r}-\boldsymbol{s}_i)=V_0\exp[{(-|\boldsymbol{r}-\boldsymbol{s}_i|)}/{r_0}]$ denotes the short-range wetting interaction, where the wetting strength of Janus nanorods on the polymer is denoted as $V_0$, and $r_0$ indicates the range of the interaction.

$F_\text{RR}$ describes the interaction free energy between Janus nanorods, expressed as follows:
\begin{equation}
F_\text{RR}=\begin{cases}
\chi\sum_{i} \sum_{j}(L-|\boldsymbol{r}_i-\boldsymbol{r}_j|)^2\cdot \big[\frac{4}{3}-\cos^2(\theta_i-\theta_j)\big],& |\boldsymbol{r}_i-\boldsymbol{r}_j|< L, \\
0,& |\boldsymbol{r}_i-\boldsymbol{r}_j|\geqslant L,
\end{cases}
\label{eq:F_RR}
\end{equation}
where $\chi$ is the interaction strength between Janus nanorods, and $L$ denotes the length of the Janus nanorods.

We perform numerical simulations of the free energy expressions (\ref{eq:F})--(\ref{eq:F_RR}) and kinetic equations~(\ref{eq:1})--(\ref{eq:4}) for the polymer nanocomposite system using cell dynamics simulations \cite{key99,key100,key101,key102} based on the CH/BD model in a $128\times128$ two-dimensional space with discretization and periodic boundary conditions. In our simulation calculations, the homopolymer component is represented by $f_\text{C}$. Furthermore, the Janus nanorods are designed as amphiphilic structures that can simultaneously wet both phase B and C, with $L_B$ representing the length of B-like sites of rod, and $NL$ denoting the number of Janus nanorods. Unless otherwise specified, the default parameters for the simulation calculations are defined as: $M_1=1.0$, $M_2=1.0$, $M_ \eta=M_\psi=1.0$, $L_\text{B}=1$, $b_2=0.2$, $\Delta t=0.6$. All parameters have been rescaled into dimensionless units.

\begin{figure}[!t]
\centerline{\includegraphics[width=0.7\textwidth]{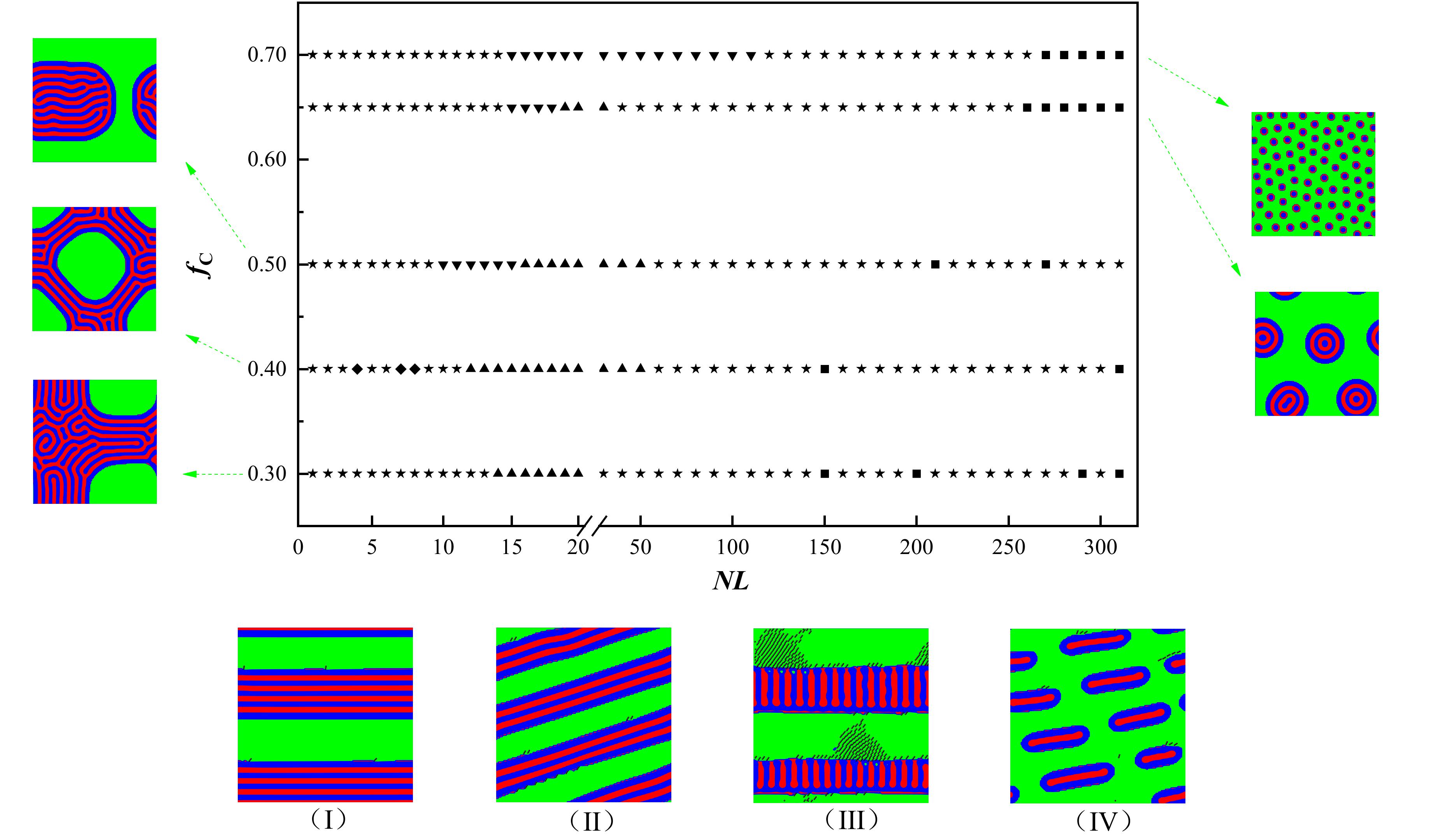}}
\caption{(Colour online) The phase diagram of symmetrical diblock copolymer/homopolymer/Janus nanorods mixed system as a function of $NL$ and $f_\text{C}$, $\psi=0$, $L=3$, $L_\text{B}=1$, $\alpha=0.02$, $b_1=0.10$, $V_0=0.08$, $\chi=0.5$. The morphologies without Janus nanorods are shown on the left-hand and right-hand sides, while the corresponding morphologies with Janus nanorods are shown below. Phase A is represented in red, phase B is represented in blue, phase C is represented in green, and the Janus nanorods are represented in black.} 
\label{fig-smp1}
\end{figure}

\section{Results and discussion}
\label{sec_3}
\subsection{Phase diagram}

The phase diagram of the symmetrical diblock copolymer/homopolymer/Janus nanorods mixed system, as a function of $NL$ and $f_\text{C}$ is presented in figure~\ref{fig-smp1}. In the absence of Janus nanorods, the polymer system undergoes a phase transition with increasing $f_\text{C}$: from a sideways ``T'' structure ($f_\text{C}=0.3$) to a roundabout structure ($f_\text{C}=0.4$), to a pie-shaped structure ($f_\text{C}=0.5$), to a concentric circular ring structure ($f_\text{C}=0.6$), and finally to a sea (phase C)-island (phase AB) structure ($f_\text{C}=0.7$). When $f_\text{C}$ is small, phase C is surrounded by the phase AB. Conversely, as $f_\text{C}$ increases and the diblock copolymer AB component gradually decreases, the phase AB transforms into the dispersed island-like structure surrounded by the phase C. With the doping of Janus nanorods into the aforementioned polymer system, the mixed system exhibits five distinct morphologies: (I) a horizontal layered structure, represented by a rhombus; (II) an tilted layered structure, represented by an equilateral triangle; (III) a vertical layered structure, represented by a square; (IV) a faulted tilted layered structure, represented by an inverted triangle. Furthermore, the disordered structure is represented by a pentagram.

\begin{figure}[!t]
\centerline{\includegraphics[width=0.85\textwidth]{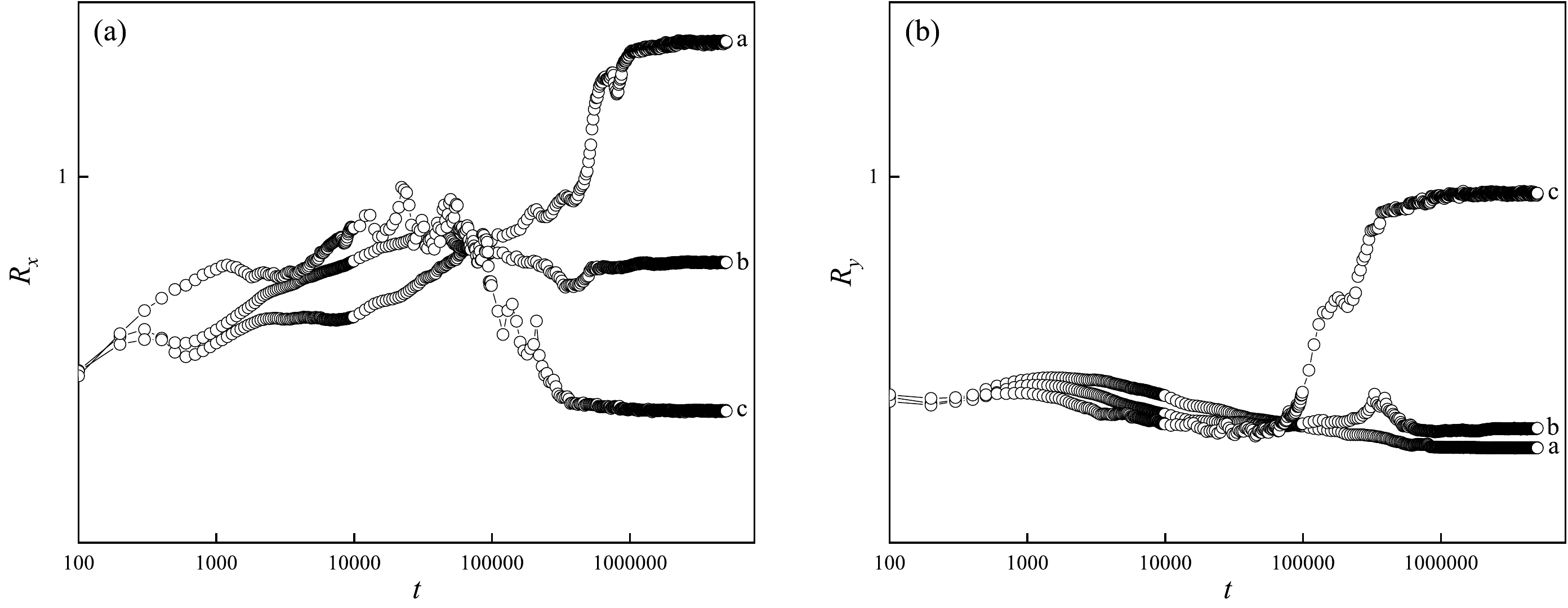}}
\caption{The double logarithmic plots of the evolution of domain sizes $R_x$ and $R_y$ in the polymer nanocomposite system as a function of time in different rod numbers at $f_\text{C}=0.40$. Curve a, $NL=4$; curve b, $NL=13$; curve c, $NL=310$.} \label{fig-smp2}
\end{figure}
\begin{figure}[!t]
\centerline{\includegraphics[width=0.85\textwidth]{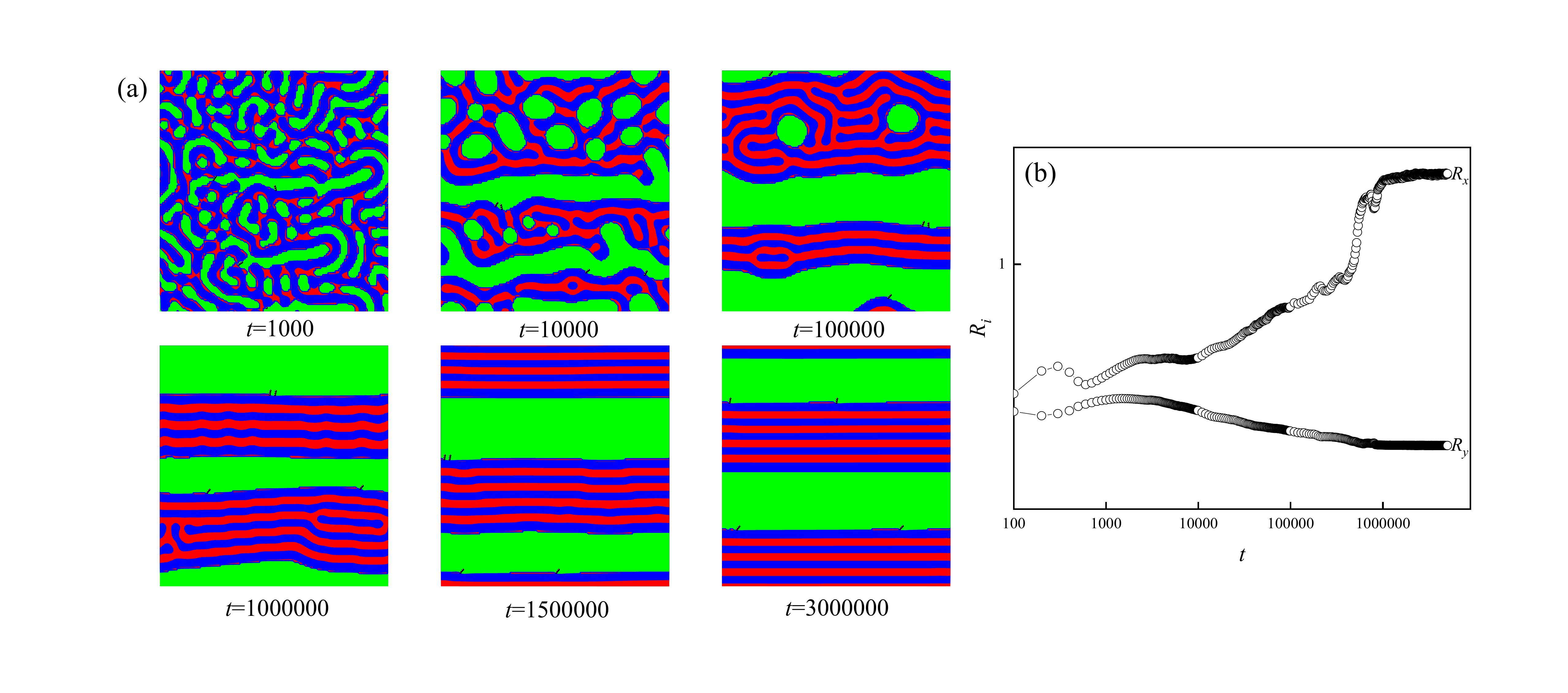}}
\caption{(Colour online) (a) Pattern evolution of the mixed system as a function of time corresponding to the horizontal layered structure in figure~\ref{fig-smp1}(I). (b) The double logarithmic plot of microdomain sizes in $x$ and $y$ direction.} \label{fig-smp3}
\end{figure}
Figure~\ref{fig-smp1} illustrates that the number of rods ranges from 1 to 310. We observe that when $f_\text{C}$ is relatively small (as shown in figure~\ref{fig-smp1} with $f_\text{C}=0.30$ and $f_\text{C}=0.40$), the polymer system exhibits a phase transition from disorder to tilted layered structures, and then to vertical layered structures with an increase of $NL$. Notably, when $f_\text{C}=0.40$ and $NL$ is low, horizontal layered structures are observed. When $f_\text{C}$ is relatively large (as shown in figure~\ref{fig-smp1} with $f_\text{C}=0.50$, $f_\text{C}=0.65$, and $f_\text{C}=0.70$), the polymer system exhibits discontinuous tilted layered structures, which can be attributed to the reduction of phase AB. As $NL$ increases, this structure eventually transforms into a continuous tilted layered structure and then into vertical layered structure. When $f_\text{C}$ is large ($f_\text{C}=0.70$), the scarcity of phase AB is pronounced, and even a large $NL$ value fails to reconnect the discontinuous tilted layered structure. 

Notably, the polymer system undergoes a transition from a tilted layered structure to a faulted tilted layered structure as $f_\text{C}$ increases within the $NL$ range of 14 to 20. In contrast to the polymer system with $f_\text{C}=0.50$, which forms a faulted tilted layered structure at $NL=10$, the polymer systems with $f_\text{C}=0.65$ and $f_\text{C}=0.70$ necessitate a substantially elevated $NL$ value of 15 for the realization of the same structural morphology. We further note that for the polymer system, the formation of a vertical layered structure occurs at $NL=270$ when $f_\text{C}=0.70$. In the case of $f_\text{C}=0.65$, the onset of the formation of this particular structure is observed at $NL=260$. When $f_\text{C}$ is 0.50, the polymer system requires $NL$ to reach 210 for the initial appearance of the vertical layered structure. Interestingly, for $f_\text{C}$ values of 0.30 and 0.40, the same vertical layered structure is formed at a relatively lower $NL$ value of 150. This indicates that the larger is $f_\text{C}$, the fewer Janus nanorods are required for the formation of a vertical layered structure. 

In this polymer system, $b_1$ is set to be 0.10, representing the repulsive interaction strength between polymers, which can be considered as the difference between the $\chi_{BC}$ and $\chi_{AC}$. Since $\chi_{BC}>\chi_{AC}$, it implies that the phase interface between the AB block copolymer and the homopolymer C should be the red A phase. However, some regions of the phase interface unexpectedly exhibit the blue B phase, as shown in the morphologies with Janus nanorods in figure~\ref{fig-smp1}. This phenomenon arises due to the presence of Janus nanorods, which are amphiphilic and simultaneously wetten both the phase B and C. As a result, the Janus nanorods induce the appearance of the B phase at certain interfacial regions.

\subsection{Evolution progress}
\begin{figure}[!t]
\centerline{\includegraphics[width=0.85\textwidth]{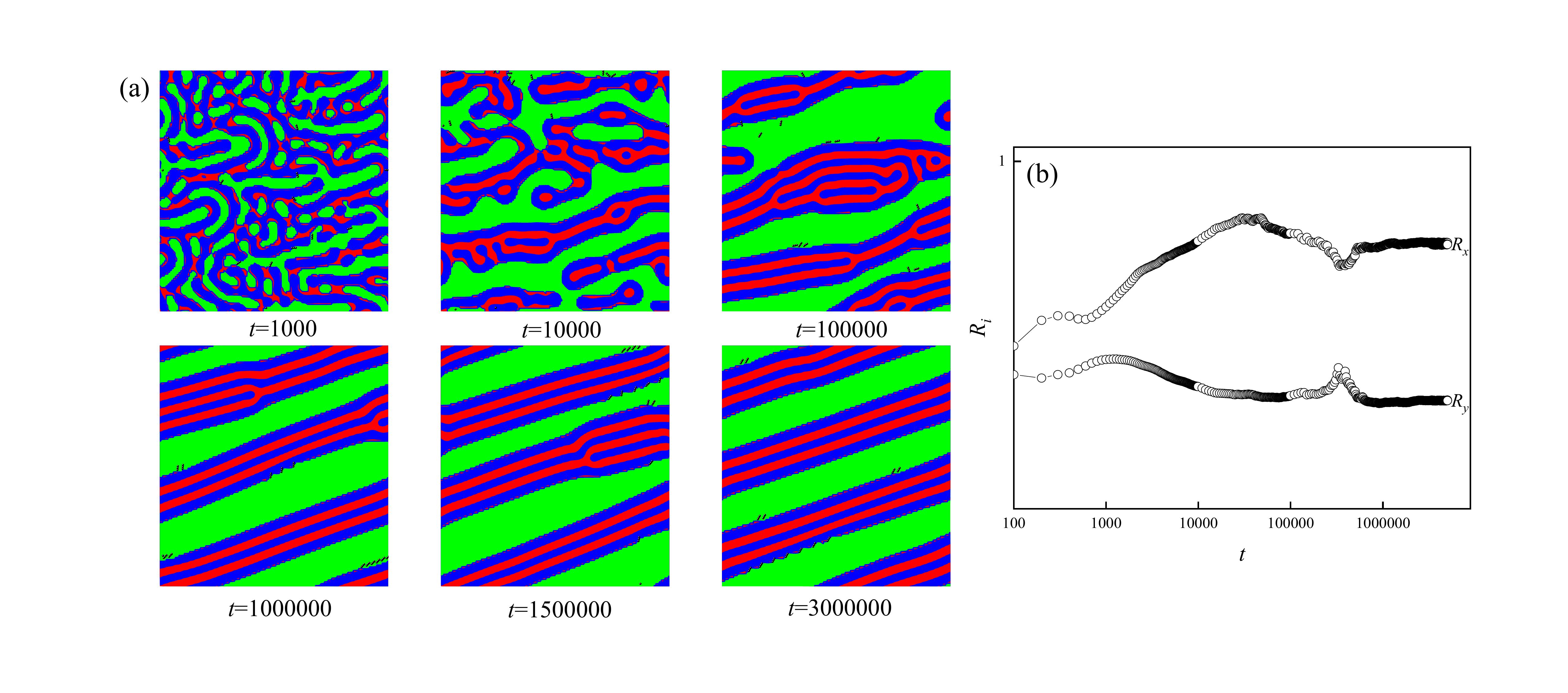}}
\caption{(Colour online) (a) Pattern evolution of the mixed system as a function of time corresponding to the horizontal layered structure in figure~\ref{fig-smp1}(II). (b) The double logarithmic plot of microdomain sizes in $x$ and $y$ direction.} \label{fig-smp4}
\end{figure}
\begin{figure}[!t]
\centerline{\includegraphics[width=0.85\textwidth]{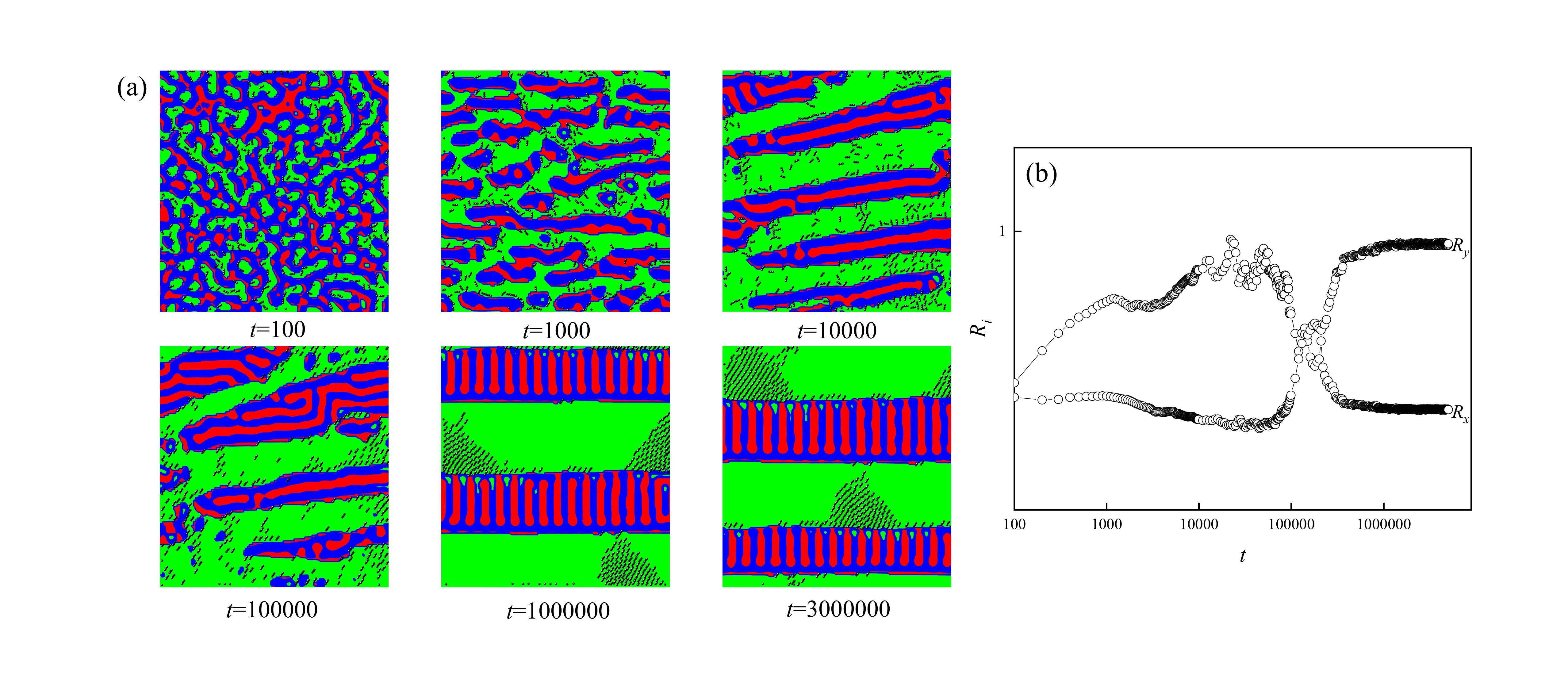}}
\caption{(Colour online) (a) Pattern evolution of the mixed system as a function of time corresponding to the horizontal layered structure in figure~\ref{fig-smp1}(III). (b) The double logarithmic plot of microdomain sizes in $x$ and $y$ direction.} \label{fig-smp5}
\end{figure}

To consider the above phenomenon in a greater depth, we numerically calculate the domain size $R_i(t)$ (where $i=x$ or $y$) as a function of time. The domain size $R_i(t)$ is defined as an inverse of the first moment of the structure factor $S(\mathbf{k},t)$:
\begin{equation}
R_i(t)=2\piup/\langle k_i(t)\rangle,
\end{equation}
where $\langle k_i(t)\rangle=\int \rd\mathbf{k} \,k_iS(\mathbf{k},t)/\int \rd\mathbf{k}\,S(\mathbf{k},t)$, the structure factor $S(\mathbf{k},t)$ depends on the Fourier components of the spatial concentration distribution \cite{key103}. Figure~\ref{fig-smp2} shows the double logarithmic plots of the microdomain size $R_i(t)$ in the $x$ and $y$ directions as a function of time for different numbers of Janus nanorods at $f_\text{C}=0.40$. Curves a, b and c correspond to figure~\ref{fig-smp1}(I), (II), and (III), with the number of Janus nanorods being $NL=4$, $NL=13$, and $NL=310$. All the results are averaged over 10 calculations. 

From figure~\ref{fig-smp2}(a), in equilibrium, the domain size $R_x(t)$ decreases as $NL$ increases (from curve a to curve c). This indicates that curve a exhibits the largest microdomain coarsening in the $x$ direction, corresponding to the horizontal layered structure in figure~\ref{fig-smp1}(I). With an increase of $NL$ (from curve a to curve b), the coarsening of microdomain in the $x$ direction is suppressed, aligning with the tilted layered structure in figure~\ref{fig-smp1}(II). The coarsening of microdomain in the $x$ direction reaches a minimum as $NL$ continues to increase (from curve b to curve c), corresponding to the vertical layered structure in figure~\ref{fig-smp1}(III).

Figure~\ref{fig-smp2}(b) shows that the domain size $R_y(t)$ in equilibrium increases as $NL$ increases (from curve a to curve c), indicating that the coarsening of microdomain in the $y$ direction intensifies with increasing $NL$. This further validates the phase transition from a horizontal layered structure to a tilted layered structure, and ultimately to a vertical layered structure. We can also see that the structure of microdomain is stable since the domain size does not change with time at the later stage.

\begin{figure}[htb]
\centerline{\includegraphics[width=0.85\textwidth]{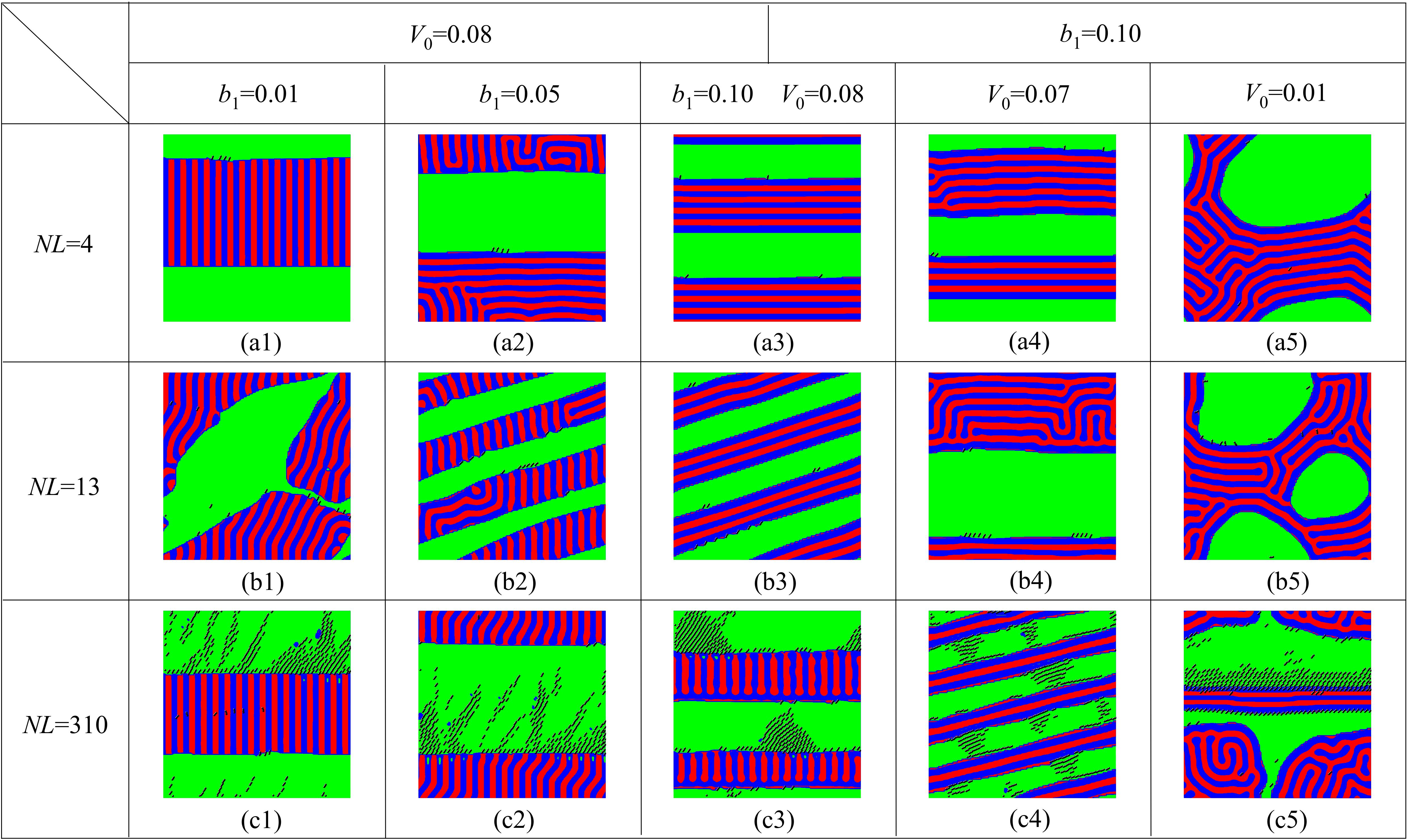}}
\caption{(Colour online) The influence of $b_1$ and $V_0$ on the morphologies of polymer nanocomposite system with different $NL$, $f_\text{C}=0.40$, $L=3$, $L_\text{B}=1$, $\alpha=0.02$, $\chi=0.5$. (a1) $b_1=0.01$, $V_0=0.08$, $NL=4$; (a2) $b_1=0.05$, $V_0=0.08$, $NL=4$; (a3) $b_1=0.10$, $V_0=0.08$, $NL=4$; (a4) $b_1=0.10$, $V_0=0.07$, $NL=4$; (a5) $b_1=0.10$, $V_0=0.01$, $NL=4$; (b1) $b_1=0.01$, $V_0=0.08$, $NL=13$; (b2)~$b_1=0.05$, $V_0=0.08$, $NL=13$; (b3) $b_1=0.10$, $V_0=0.08$, $NL=13$; (b4) $b_1=0.10$, $V_0=0.07$, $NL=13$; (b5) $b_1=0.10$, $V_0=0.01$, $NL=13$; (c1) $b_1=0.01$, $V_0=0.08$, $NL=310$; (c2)~$b_1=0.05$, $V_0=0.08$, $NL=310$; (c3) $b_1=0.10$, $V_0=0.08$, $NL=310$; (c4) $b_1=0.10$, $V_0=0.07$, $NL=310$; (c5)~$b_1=0.10$, $V_0=0.01$, $NL=310$. Phase A is represented in red, phase B is represented in blue, phase C is represented in green, and the Janus nanorods are represented in black.} \label{fig-smp6}
\end{figure}

To monitor the process of forming the ordered structures corresponding to figure~\ref{fig-smp1}(I), (II), and (III), we present the pattern evolution of the mixed system for $NL$ values of 4, 13, and 310, as shown in figures~\ref{fig-smp3}, \ref{fig-smp4}, and \ref{fig-smp5}. Meanwhile, we also provide a comparison between the microdomain sizes in the $x$ and $y$ directions. 

\begin{figure}[htb]
\centerline{\includegraphics[width=0.8\textwidth]{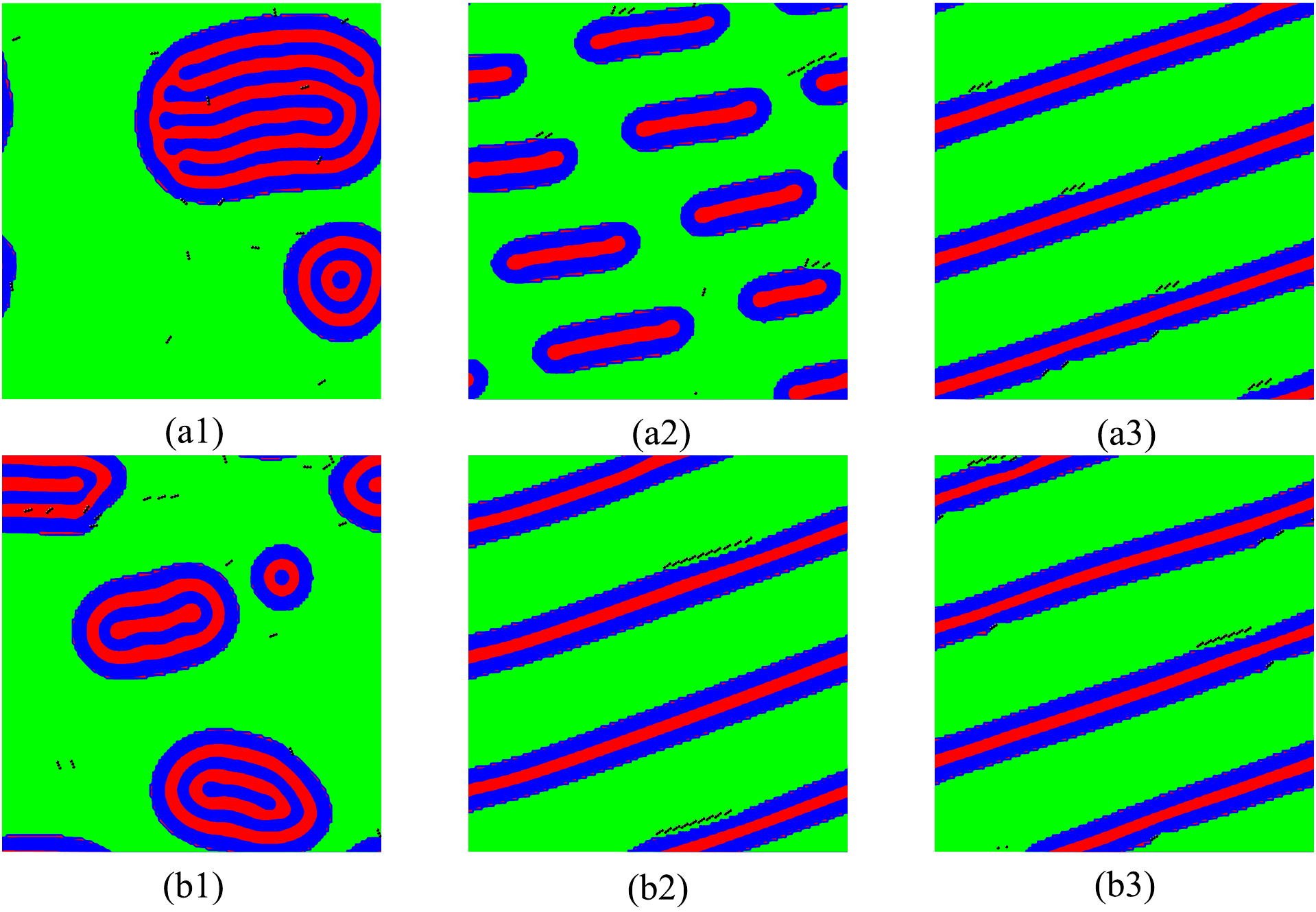}}
\caption{(Colour online) The influence of $V_0$ on the morphologies of polymer nanocomposite system with different $NL$, $f_\text{C}=0.65$, $L=3$, $L_\text{B}=1$, $\alpha=0.02$, $b_1=0.10$, $\chi=0.5$. (a1) $V_0=0.01$, $NL=15$; (a2) $V_0=0.08$, $NL=15$; (a3) $V_0=0.12$, $NL=15$; (b1) $V_0=0.01$, $NL=19$; (b2) $V_0=0.08$, $NL=19$; (b3) $V_0=0.12$, $NL=19$. Phase A is represented in red, phase B is represented in blue, phase C is represented in green, and the Janus nanorods are represented in black.} \label{fig-smp7}
\end{figure}

\begin{figure}[htb]
\centerline{\includegraphics[width=0.8\textwidth]{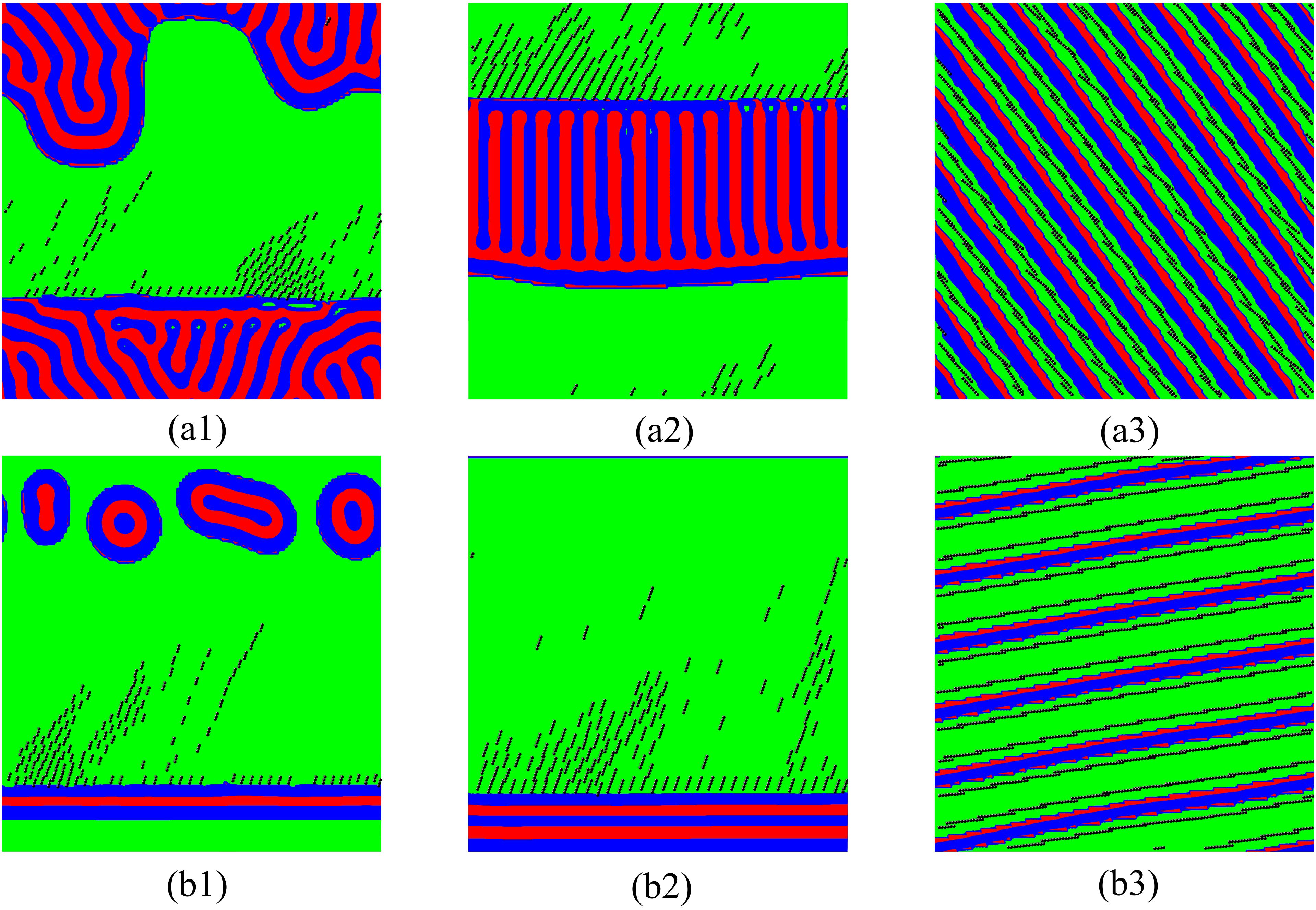}}
\caption{(Colour online) The influence of $L$ on the morphologies of polymer nanocomposite system with different $f_\text{C}$, $L_\text{B}=1$, $NL=140$, $\alpha=0.02$, $b_1=0.10$, $V_0=0.08$, $\chi=0.5$. (a1) $f_\text{C}=0.40$, $L=3$; (a2)~$f_\text{C}=0.40$, $L=4$; (a3) $f_\text{C}=0.40$, $L=12$; (b1) $f_\text{C}=0.70$, $L=3$; (b2) $f_\text{C}=0.70$, $L=4$; (b3)~$f_\text{C}=0.70$, $L=12$. Phase A is represented in red, phase B is represented in blue, phase C is represented in green, and the Janus nanorods are represented in black.} \label{fig-smp8}
\end{figure}

Figure~\ref{fig-smp3} shows that the polymer system is in a disordered state in the early stage of phase separation ($t=1000$), with Janus nanorods distributed randomly at the phase interface. Correspondingly, the domain size $R_x$ exceeds $R_y$ to a small extent, and they are approximately equal. During the middle stage of phase separation ($1000<t<1000000$), the polymer system begins to exhibit a tendency towards the formation of a horizontal layered structure. In the late stage of phase separation ($t\geqslant1000000$), the polymer system exhibits an ordered horizontal layered structure, with the Janus nanorods anchored nearly perpendicular to the phase interface between phases B and C. Moreover, it is evident that the domain size $R_x$ is substantially greater than $R_y$ at equilibrium, providing further confirmation of the previously observed parallel layered structure.

\subsection{The influence of other parameters}
\begin{figure}[!t]
\centerline{\includegraphics[width=0.70\textwidth]{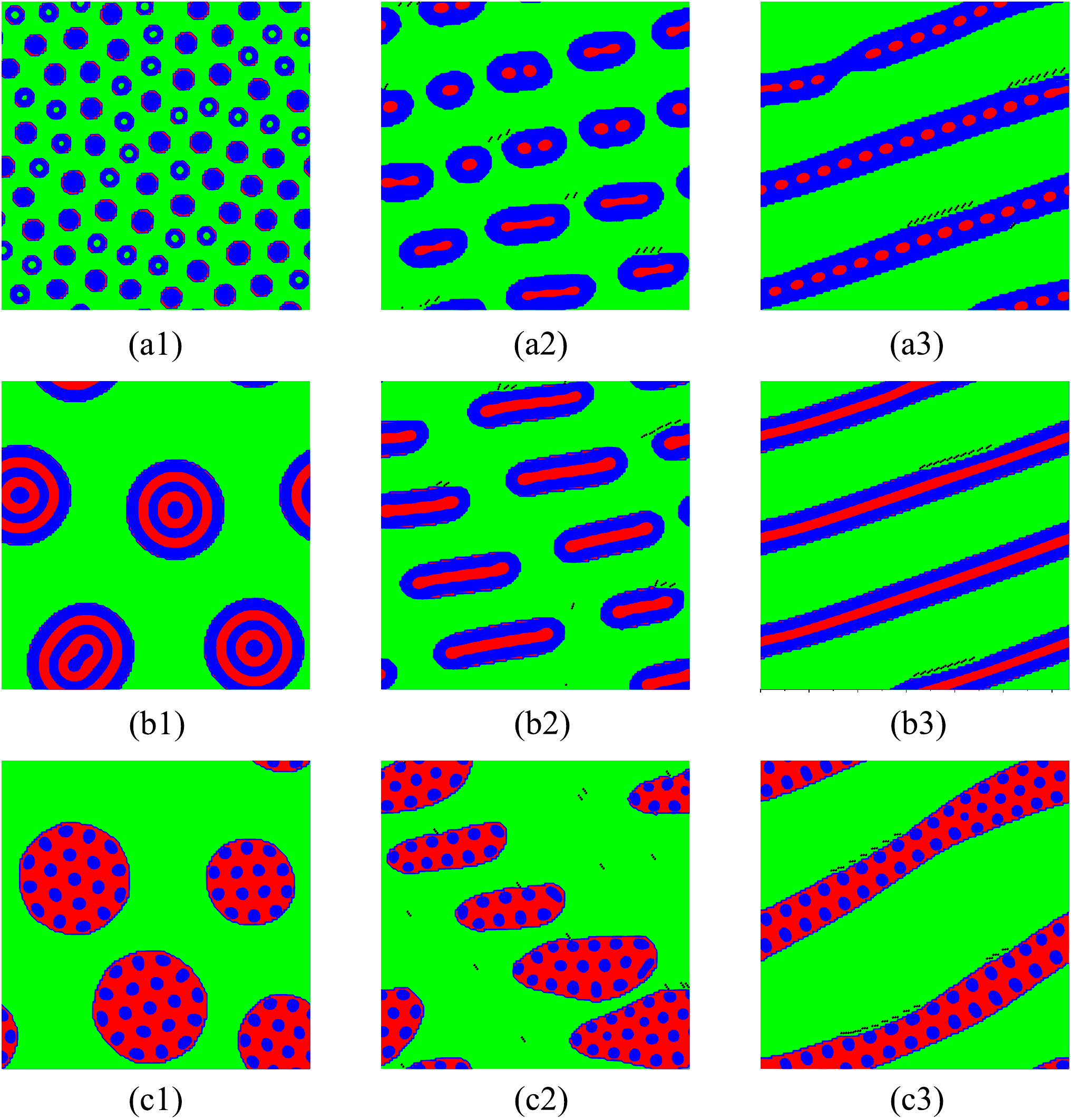}}
\caption{(Colour online) The influence of $\psi$ on the morphologies of polymer nanocomposite system with different $NL$, $f_\text{C}=0.65$, $L=3$, $L_\text{B}=1$, $\alpha=0.02$, $b_1=0.10$, $V_0=0.08$, $\chi=0.5$. (a1) $\psi=-0.25$, $NL=0$; (a2) $\psi=-0.25$, $NL=15$; (a3) $\psi=-0.25$, $NL=19$; (b1) $\psi=0$, $NL=0$; (b2) $\psi=0$, $NL=15$; (b3) $\psi=0$, $NL=19$; (c1) $\psi=+0.25$, $NL=0$; (c2) $\psi=+0.25$, $NL=15$; (c3) $\psi=+0.25$, $NL=19$. Phase A is represented in red, phase B is represented in blue, phase C is represented in green, and the Janus nanorods are represented in black.} 
\label{fig-smp9}
\end{figure}

When the number of Janus nanorods is 13, the resulting morphology exhibits a tilted layered structure. The corresponding morphological evolution and the domain sizes $R_i$ are illustrated in figure~\ref{fig-smp4}. Similar to figure~\ref{fig-smp3}, at the early stage of phase separation ($t=1000$), the polymer system also shows a disordered structure. However, at this stage, $R_x$ is slightly larger than $R_y$, indicating a tendency for the structure to transition towards a tilted layered configuration. Subsequently, the polymer system gradually transitions to a tilted layered structure as time evolves, with the Janus nonorods being pinned at the phase interface at a certain angle. The variations of $R_x$ and $R_y$ have slowed considerably compared to those observed in figure~\ref{fig-smp3}. However, $R_x$ ultimately remains greater than $R_y$, further substantiating that the rusulting tilted layered structure is oriented preferentially along the $x-$ direction.

As the number of Janus nanorods increases considerably to 310, the polymer system ultimately evolves into a vertical layered structure. The corresponding morphological evolution and the growth curve of the domain size are illustrated in figure~\ref{fig-smp5}. Similarly, the disordered structure corresonds to $R_x\approx R_y$. As time progresses, the polymer system first forms a defective tilted layered structure ($t=10000$), at which point $R_x$ is considerably greater than $R_y$. As time continues to increase, the polymer system gradually transitions to a vertical layered structure. Thus, the domain size $R_x$ steadily decreases, while $R_y$ gradually increases, until it reaches a stable state where $R_y$ is considerably greater than $R_x$, corresponding to the vertical layered structure.

Systematic evaluations of the effects of various parameters on the polymer system provide a valuable guidance for the assembly of the nanocomposites. In addition to the number of nanorods and the composition ratio, we also consider the effects of the repulsive interaction strength between polymers, the wetting strength, the length of nanorods, and the degree of asymmetry.

\subsubsection{The repulsive interaction strength and wetting strength}
Figure~\ref{fig-smp6} shows the influence of the wetting strength $V_0$ and the repulsive interaction strength between polymers $b_1$ on the phase behavior of the polymer system. As shown in figure~\ref{fig-smp6}, the morphologies in the first, second, and third rows correspond to different numbers of Janus nanorods ($NL$): 4, 13, and 310, respectively. The morphologies in the central column correspond to $b_1=0.10$ and $V_0=0.08$, while the two columns on the left correspond to $b_1$ values of 0.01 and 0.05, and the two columns on the right correspond to $V_0$ values of 0.07 and 0.01, respectively.

We observe that the polymer system undergoes a phase transition from parallel lamellar structures to tilted lamellar structures and eventually to perpendicular lamellar structures as the number of nanorods increases, a phenomenon also noted in the phase diagram presented in figure~\ref{fig-smp1}. However, if either $b_1$ or $V_0$ decreases, different structures will emerge. When the number of nanorods is small ($NL=4$), the polymer system gradually transitions from the original parallel lamellae [figure~\ref{fig-smp6}(a3)] to perpendicular lamellae [figure~\ref{fig-smp6}(a1)] as $b_1$ decreases. For moderate values of $b_1$, the system exhibits a coexistence structure of parallel and perpendicular orientations [figure~\ref{fig-smp6}(a2)]. The underlying reason is that when $b_1$ is large, the repulsive interaction between phase B and C is much stronger than that between phase A and C, leading to a lamellar structure dominated by the red A phase at the phase interface. However, since the Janus nanorods simultaneously wet both phases B and C, their effect results in the appearance of the blue B phase at certain interfacial regions. As $b_1$ decreases further, the difference between $\chi_{BC}$ and $\chi_{AC}$ diminishes. An increased interfacial contact area between the phases A and C leads to the formation of a coexistence structure of parallel and perpendicular orientations. When the number of nanorods is moderate ($NL=13$), the polymer system transitions from a slanted layered structure with parallel alignment of the AB phase [figure~\ref{fig-smp6}(b3)] to a slanted layered structure where the AB phase coexists in both parallel and perpendicular orientations [figure~\ref{fig-smp6}(b2)] as $b_1$ decreases. As $b_1$ decreases further to 0.01, the polymer system transitions into a disordered state, with the interfacial contact area between phases A and C being approximately equal to that between phases B and C. However, when the number of Janus nanorods is relatively high ($NL=310$), the polymer system has already formed a perpendicular layered structure induced by the Janus nanorods. Thus, a decreasing $b_1$ has an effect similar to the increasing number of nanorods. Consequently, the decrease of $b_1$ does not significantly impact the polymer system, which remains in a perpendicular layered structure. 

When decreasing the wetting strength $V_0$, it is evident that the interaction between the rods and the phase is weakened, which implies that the effect of the Janus nanorods on the phase behavior of the polymer system is also diminished. A larger rod number and a stronger rod-phase interaction result in the formation of a perpendicular layered structure with $V_0=0.08$ and $NL=310$ [figure~\ref{fig-smp6}(c3)]. Nevertheless, if $V_0$ decreases, a similar tilted layere structure appears [figure~\ref{fig-smp6}(c4)], which is equivalent to the effect of reducing the number of rods. A further decrease in $V_0$ causes the polymer system to transform into a disordered state [figure~\ref{fig-smp6}(c5)]. In other cases, the effects of reducing the number of nanorods and decreasing the wetting strength on the phase behavior of the polymer system are comparable. This conclusion is particularly evident when $f_\text{C}=0.65$, as shown in figure~\ref{fig-smp7}.

Overall, as the repulsive interaction strength $b_1$ decreases and the number of nanorods $NL$ increases, the polymer system gradually transitions from a parallel layered structure to a tilted layered structure, ultimately reaching a perpendicular layered structure. Furthermore, the changes in the wetting strength $V_0$ also influence this transition. As the wetting strength $V_0$ decreases, the interaction between the nanorods and the phase weakens, which results in a structural change trend similar to that observed when the number of nanorods is reduced.

In figure~\ref{fig-smp7}, the corresponding number of Janus nanorods for (a$i$) and (b$i$) is 15 and 19, respectively, with each column corresponding to the wetting strength of 0.01, 0.08, and 0.12. In the absence of nanorods, the polymer system exhibits a regular concentric ring structure, as depicted in the morphology in figure~\ref{fig-smp1}. Upon doping with Janus nanorods, both the wetting strength and the number of nanorods influence the final morphological structure. At a low wetting strength, specifically at 0.01, the effect of the nanorods is minimal, resulting in the formation of an elongated concentric ring structure of the polymer system. When $NL=15$ and the wetting strength increases to 0.08, it forms a faulted tilted layered structure. Further increasing the wetting strength to 0.12 leads to the formation of a regular tilted layered structure in the blend system. However, when the number of nanorods increases to 19, a tilted layered structure forms even at the lower wetting strength of 0.08. This indicates that increasing the number of nanorods has a similar effect on the phase behavior of the polymer system as increasing the wetting strength. This conclusion is consistent with the findings presented in figure~\ref{fig-smp6}.

\subsubsection{The length of nanorods}
Additionally, we explore the effect of the length of nanorods on the polymer system. Figure~\ref{fig-smp8} shows the example of morphologies for the self-assembly of polymer system with different rod lengths. At $f_c=0.4$ and $NL=140$, when the length of the Janus nanorods is 3, the polymer system exhibits a disordered structure [figure~\ref{fig-smp8}(a1)]. Increasing $L$ to 4 causes it to transition to a horizontally oriented comb-like structure [figure~\ref{fig-smp8}(a2)]. Both in the disordered structure and in the horizontally oriented comb-like structure, the Janus nanorods are pinned at the phase interface at a certain angle. However, due to the excessive number of nanorods, some of them become detached from the phase interface and disperse within phase C. When the rod length is further increased to 12, while the number of nanorods remains constant, the increased rod length enhances the rod-phase interaction, causing the polymer system to transform into a tilted layered structure [figure~\ref{fig-smp8}(a3)]. The Janus nanorods are too long for the spacing between two layers to accommodate them; thus, they exhibit an end-to-end arrangement in a nanowire-like structure that is nearly parallel to the phase interface.When $f_C$ is increased to 0.70, maintaining the same number of nanorods, the polymer system transitions from a structure with coexisting concentric rings and parallel layers [figure~\ref{fig-smp8}(b1)] to a parallel layered structure [figure~\ref{fig-smp8}(b2)]. Finally, when the nanorods are long, the system forms a tilted layered structure [figure~\ref{fig-smp8}(b3)]. At this moment, the end-to-end nanowire-like structure that is parallel to the phase interface becomes more orderly.

\subsubsection{The degree of asymmetry}

In all the previously described polymer systems, the diblock copolymer is symmetric, i.e., $\psi=0$. However, if the diblock copolymer is asymmetric, the polymer system exhibits different phase transitions. For instance, the polymer system with symmetric diblock copolymer displays a well-ordered concentric ring structure [figure~{\ref{fig-smp9}}(b1)] in the absence of nanorods at $f_C=0.65$. After doping the Janus nanorods, the system gradually transitions to a faulted tilted layered structure [figure~\ref{fig-smp9}(b2)] as the number of nanorods increases, eventually forming an ordered tilted layered structure [figure~\ref{fig-smp9}(b3)]. However, when $\psi=-0.25$, in the absence of nanorods, the polymer system forms a dispersed sea-island structure [figure~{\ref{fig-smp9}}(a1)], where phase C is the sea and phase AB forms concentric ring islands. Due to the scarcity of the A component, the concentric ring island structure contains a very small proportion of the red A phase. The introduction of Janus nanorods induces the aggregation of the dispersed island, resulting in an elongated sea-island structure [figure~\ref{fig-smp9}(a2)]. Thereby on account of the decrease of phase A, the phase A within the islands is not fully interconnected and some portions of it appear as point-like structure. As the number of nanorods is increased to 19, the elongated sea-island structures are connected, forming a tilted layered structure overall, where the AB phase takes on a pod-like configuration. When $\psi=+0.25$, the A component is considerably larger than the B component, leading to the formation of a lotus root-like structure in the absence of nanorods [figure~\ref{fig-smp9}(c1)]. In the case of the number of Janus nanorods being 15, the lotus root-like structure is similarly elongated [figure~\ref{fig-smp9}(c2)]. However, under the effect of a relatively larger number of Janus nanorods, the polymer system also displays a tilted layered configuration, where the AB phase also presents a pod-like structure, with the beans now being phase B [figure~\ref{fig-smp9}(c3)].

\section{Conclusions}
\label{sec_4}

We use the cell dynamics simulation based on CH/BD model to investigate the self-assembly behavior of the mixed system consisting of diblock copolymers (AB)/homopolymers (C)/Janus nanorods. The results show that when $f_\text{C}$ is relatively small, the polymer system exhibits a phase transition from disordered to tilted layered, and then to perpendicular layered structures with the increase of $NL$. Notably, when $f_\text{C}=0.40$ and $NL$ is in the lower range, horizontal layered structures are observed. When $f_\text{C}$ is relatively large, the polymer system presents discontinuous tilted layered structures due to the reduction of phase AB, which eventually transforms into a continuous tilted layered structure and vertical layered structure as $NL$ increases. However, when $f_\text{C}$ is large ($f_\text{C}=0.70$), the scarcity of phase AB is so pronounced that, despite a substantially large $NL$, it fails to prompt the reconnection of the discontinuous tilted layered structure in the polymer system. To validate this phenomenon in greater depth, we examined the dynamic evolution of domain size and the pattern evolution as a function of time.

Additionally, the repulsive interaction strength between polymers, the wetting strength, the length of nanorods, and the degree of asymmetry significantly influence the phase separation of the polymer system. The effect of $b_1$ on the blend system differs depending on the number of nanorods. With a small number of Janus nanorods, the polymer system gradually transitions from the original parallel lamellae to a coexistence structure of parallel and perpendicular orientationsand then to perpendicular lamellae as $b_1$ decreases. With a moderate number of Janus nanorods, the system transitions from a slanted layered structure with parallel alignment of the AB phase to one where the AB phase coexists in both parallel and perpendicular orientations, and then into a disordered state. However, when the number of nanorods is very high, decreasing $b_1$ does not considerably impact the polymer system, which remains in a perpendicular layered structure. Furthermore, increasing the wetting strength has a similar effect on the phase behavior of the polymer system as increasing the number of nanorods. For the length of nanorods, increasing $L$ favors the emergence of a tilted layered structure, while the Janus nanorods form the nanowire structure that is nearly parallel to the phase interface. In addition, if the diblock copolymer is asymmetric, the polymer system exhibits different transitions with variations in the number of nanorods at different degrees of asymmetry. These results provide both theoretical and experimental references for the preparation of new types of high-performance nanomaterials.

\section*{Acknowledgements}
I would like to thank all the authors of this article for their contributions, with Y.Q. Guo and J. Liu having contributed equally to this work. This research was supported by the National Natural Science Foundation of China (Grant Nos. 22403059 and 22378411) and the Basic Research Program of Shanxi Province (Grant No. 202103021223386).

\ukrainianpart

\title{Поведінка самозбірки диблокового кополімеру/гомополімеру, індукована наностержнями Януса}
\author{І. К. Го\refaddr{label1}, Дж. Лю\refaddr{label2,label3}, Х. Р. Хе\refaddr{label4}, Н. Ву\refaddr{label1}, Дж. Дж. Чжан\refaddr{label2,label3}}
\addresses{
	\addr{label1} Факультет хімічної інженерії та матеріалознавства, Університет Люлян, 033001 Ліші, Китай
	\addr{label2} Школа фізики та інформаційної інженерії, Шаньсійський педагогічний університет, 030031 Тайюань, Китай
	\addr{label3}Шаньсійський інститут вуглехімії, Китайська академія наук, 033001 Тайюань, Китай
	\addr{label4} Школа хімії та хімічної інженерії, Шаньсійський університет, 030006 Тайюань, Китай
}

%
%
%

\makeukrtitle

\begin{abstract}
	\tolerance=3000%
	Ми використовуємо моделювання динаміки комірок на основі моделі CH/BD для дослідження поведінки самозбірки змішаної системи, що складається з диблок-кополімерів (AB), гомополімерів (C) та наностержнів Януса. Результати показують, що при різних співвідношеннях компонентів змішана система зі збільшенням кількості наностержнів зазнає різних фазових переходів. Зокрема, коли концентрація гомополімерного компонента становить 0.40, змішана система переходить від невпорядкованої структури до паралельної пластинчастої структури, потім до похилої шаруватої структури і, зрештою, до перпендикулярної пластинчастої структури зі збільшенням кількості наностержнів. Щоб глибше дослідити це явище, ми проводимо комплексний аналіз розмірів доменів та еволюції структури. Крім того, ми досліджуємо вплив сили відштовхувальної взаємодії між полімерами, змочування, довжини наностержнів та ступеня асиметрії на поведінку самозбірки змішаної системи. Це дослідження дає значну теоретичну та експериментальну інформацію для отримання нових наноматеріалів.
	\keywords самозбірка, диблок-сополімер, гомополімер, наностержні Януса
	
\end{abstract}

\lastpage
\end{document}